\documentclass[12pt]{article}

\newtheorem{Theorem}{Theorem}[section]

\newtheorem{Proposition}[Theorem]{Proposition}

\makeatletter
\usepackage{latexsym}
\usepackage[T1]{fontenc}
\usepackage{amsmath}
\usepackage{amssymb}

\def \AO {{\cal A}({\cal O})}
\def \AO' {{\cal A}({\cal O}')}

\def \] {\supseteq}

\def \Pf {{\bf Proof.\,\,}}

\def \limR {\lim_{R \ra \infty}}

\def \be {\begin{equation}}
\def \be {\begin{equation}}
\def \ee {\end{equation}}
\def \ume {{\scriptstyle{\frac{1}{2}}}}
\def \ra {\rightarrow}

\def \eqq {\equiv}

\def \a {{\alpha}}
\def \b {{\beta}}
\def \g {{\gamma}}
\def \d {{\delta}}
\def \eps {{\varepsilon}}

\def \l {{\lambda}}

\def \s {{\sigma}}

\def \ph {{\varphi}}

\def \o {{\omega}}

\def \A {{\cal A}}
\def \B {{\cal B}}
\def \C {{\cal C}}
\def \D {{\cal D}}
\def \F {{\cal F}}
\def \G {{\cal G}}
\def \H {\mbox{${\cal H}$}}

\def \L {{\cal L}}

\def \O {{\cal O}}

\def \S {{\cal S}}

\def \Z {{\cal Z}}

\def \Psio {{\Psi_0}}

\def \do {{\partial_0}}

\def \dmu {{\partial_\mu}}
\def \dnu {{\partial_\nu}}
\def \dla {{\partial_\lambda}}

\def \dum {{\partial^\mu}}
\def \dun {{\partial^\nu}}

\def \Fmn {{F_{\mu\,\nu}}}

\def \k {{\bf k}}

\def \j   {{\bf j}}

\def \q {{\bf q}}

\def \x {{\bf x}}
\def \y {{\bf y}}

\def \z {{\bf z}}

\def \Rbf {{\bf R}}

\def \AO {{\cal A}({\cal O})}
\def \AO' {{\cal A}({\cal O}')}

\def \] {\supseteq}

\def \Pf {{\bf Proof.\,\,}}

\def \limR {\lim_{R \ra \infty}}

\newfam\Ssfam
\font\eleSs=cmss10 at12pt \font\sevenSs= cmss10 at 8pt \font\sixSs= cmss10 at
6pt

\textfont\Ssfam=\eleSs\scriptfont\Ssfam=\sevenSs%
\scriptscriptfont\Ssfam=\sixSs \def\Ss{\fam\Ssfam\eleSs}

\def\doppio#1{{\rm I}\kern-.1667em{\rm #1}}

\def\Q{\text{Q}\kern-.52em
    \text{\vrule height1.5ex width.5pt depth0pt}\kern.45em}

\def\Z{{\mathchoice {\hbox{$\Ss\textstyle Z\kern-0.4em Z$}}
{\hbox{$\Ss\textstyle Z\kern-0.4em Z$}} {\hbox{$\Ss\scriptstyle Z\kern-0.25em
Z$}} {\hbox{$\Ss\scriptscriptstyle Z\kern-0.2em Z$}}}}

\def\C{{\mathchoice{\hbox{$\rm\textstyle\text{\kern.35em\vrule
   height1.5ex width.5pt depth0pt\kern-.35em C}$}}
{\hbox{$\rm\textstyle\text{\kern.35em\vrule
   height1.5ex width.5pt depth0pt\kern-.35em C}$}}
{\hbox{$\rm\scriptstyle\text{\kern.35em\vrule
   height1.5ex width.3pt depth0pt\kern-.35em C}$}}
{\hbox{$\rm\scriptscriptstyle\text{\kern.35em\vrule
   height1.5ex width.2pt depth0pt\kern-.35em C}$}}}}

\def \be{\begin{equation} \displaystyle}
\def \ee{\end{equation}}

\def \A*{\mbox{$A^{*} $}}
\def \B*{\mbox{$B^{*} $}}
\def \C*{\mbox{$C^{*} $}}

\def \bea{\begin{eqnarray}}
\def \eea{\end{eqnarray}}

\def \Pf{{\em Proof.\,\,}}

\def \a{\alpha}
\def \b{\beta}
\def \g{\gamma}
\def \l{\lambda}
\def \d{\delta}

 \@addtoreset{equation}{section}

\def \be {\begin{equation} \displaystyle}
\def \ee {\end{equation}}
\def \ra {\rightarrow}
\def\AO {\mbox{${\cal A}({\cal O})$}}
\def\AO'{\mbox{${\cal A}({\cal O}')$}}
\def\O {\mbox{${\cal O}$}}

\def\A{\mbox{${\cal A}$}}

\def \ra{\rightarrow}
\def \ph {{\varphi}}
\def \eps {{\varepsilon}}

\def \o {{\omega}}
\def \O {{\cal O}}

\def \A {{\cal A}}
\def \AO {\A(\O)}
\def \AOl'{\A(\O_{loc}')}

\def \B {{\cal B}}
\def \F {{\cal F}}
\def \D {{\cal D}}

\def \H {{\cal H}}

\def \S {{\cal S}}

\def \Fmn {{F_{\mu \nu}}}

\def \dmu {{\partial^\mu}}
\def \dnu {{\partial^\nu}}

\def \k {{\bf k}}
\def \x {{\bf x}}

\def \z {{\bf z}}

\def \y {{\bf y}}

\def \g {{\gamma}}

\begin{document}
\title{Localization and symmetries}
\sloppy
\author{G. Morchio \\Dipartimento di Fisica, Universita' di Pisa, \\and  INFN, Sezione di Pisa
\and F. Strocchi \\ Scuola Normale Superiore, and INFN, Sezione  di Pisa}

\fussy
\date{} \maketitle
\begin{abstract}
The violation of the Noether relation between symmetries and charges is reduced
to the time dependence  of the charge associated  to a conserved current. For
the $U(1)$ gauge symmetry a non-perturbative control of the charge commutators
is obtained by an analysis of the Coulomb charged fields. From this, in the
unbroken case  we obtain a correct expression for the electric charge on the
Coulomb states, its superselection and the presence of massless vector bosons;
in the broken case, we obtain a general non-perturbative version of the Higgs
phenomenon, i.e. the absence of massless Goldstone bosons and of massless
vector bosons. The conservation of the (gauge dependent) current associated to
the $U(1)$ axial symmetry in QCD is shown to be compatible with the time
dependence of the corresponding charge commutators and a non-vanishing $\eta'$
mass, as a consequence of the non locality of the (conserved) current.

\end{abstract}

\maketitle
\section{Introduction}
The role of continuous symmetries and their breaking in the recent developments
of theoretical physics needs not to be further discussed here. However, in our
opinion, the interplay between symmetries and localization properties of the
fields for  infinitely extended systems is not sufficiently emphasized in the
textbook presentations, in particular in connection with the phenomenon of
symmetry breaking in gauge theories and in Coulomb systems.

We shall focus our attention to the case of  continuous symmetries which
commute with space translations and with time evolution. For infinitely
extended systems described by field variables, $\ph(\x, t)$, and by  a
Lagrangean function,  an {\em internal symmetry} is a transformation of the
fields $g :  \ph(\x, t) \ra (g \ph)(\x, t)$, $g$ independent of $\x$ and $t$,
which leaves the Lagrangean density invariant.  At the level of local variables
and measurements the implications of such an invariance property is not as
direct as it appears. In fact, the invariance of the Lagrangean under a
(continuous) one parameter group of symmetries implies the existence of a {\em
conserved current} $j_\mu(\x, t)$, \be{\partial_t j_0 + \mbox{div} \j \eqq \dum
j_\mu (x) =0,}\ee i.e. a {\em local conservation law}. However, the
implications of such a local conservation, in particular the existence of a
conserved charge or the existence of Goldstone bosons in the case of symmetry
breaking, critically depend on the localization properties of the relevant
variables and configurations.

This problem has been extensively discussed in the literature~\cite{KRS, SWIE1}
under the assumption that, according to the general wisdom of  Noether theorem,
the time independent symmetry transformations of the fields are generated by
the space integral of the charge density $j_0$ of the corresponding Noether
conserved current $j_{\mu}$: \be{ \d A = i \lim_{R \ra \infty} [\,Q_R,
\,A\,],}\ee $$Q_R \eqq j_0(f_R, \a) \eqq \int d^{s+1}x j_0(\x, t) f_R(\x)
\a(x_0),$$ where the smearing test functions $f_R(\x) = f(|\x|/R)$, $f, \,\a\,
\in {\cal D}(\Rbf)$, $f(x) = 1 $, for $|x| \leq R$, take care of the necessary
ultraviolet regularization. It is enough that the limit exists for the field
correlation functions. The independence of the r.h.s of eq.\,(1.2) from the
choice of the test function $\a$, with the normalization condition $\int d
x_0\,\a(x_0) =1$, is formally equivalent to the time independence of the
(space) integral of the charge density and  therefore it is necessary for the
validity of eq.\,(1.2). It is also assumed that eq.\,(1.2) holds independently
of whether the symmetry is broken or not.

The validity of such assumptions follows if $j_\mu$ and $A$ are {\em relatively
local}, e.g. if the canonical structure is local and the time evolution of both
$j_\mu$ and $A$ is relativistically causal; in this case, the limit is reached
for finite values of $R$ and eq.\,(1.1) implies the independence of $\a$ and
eq.\,(1.2). The same conclusion holds if the  delocalization induced by the
time evolution is not worse than $r^{-2 -\eps}$, $ \eps > 0$. However,
important physical phenomena are governed by time independent symmetries for
which the current conservation, eq.\.(1.1), does not imply the generation of
the symmetry  by the integral of the charge density, eq.\,(1.2), and one cannot
rely on the above assumptions.

The crucial issue is the time dependence of the integral of the charge density,
namely the $\a$ dependence of the r.h.s of eq.\,(1.2), in spite of the
conservation of the current, i.e. the failure of sufficient relative locality
between $j_i$ and the operator $A$. The aim of this note is to critically
examine the mechanisms at the basis of such a failure and their physical
consequences both in the case of an exact and of a broken symmetry.

The inevitable non locality of the charged fields has been proved to follow if
the current obeys a local Gauss law: $j_\mu = \partial^\nu F_{\nu\, \mu}$, with
$\Fmn = - F_{\nu \,\mu}$ a local field~\cite{FPS}; however, this does not
directly imply that the charged fields are not local with respect to $j_i$
(e.g. in the classical Maxwell-Dirac and Maxwell-Klein-Gordon the charged
fields are local with respect to $j_i$~\cite{BDM}) and, even more importantly,
that there is enough relative non-locality to force the $\a$ dependence of the
r.h.s of eq.\,(1.2) and, therefore, the  {\em violation the Noether relation
between the symmetry and the integral of the charge density of the
corresponding current}, eq.\,(1.2). In Section 3, by applying the analysis of
Refs.\,~\cite{BDM, MS} we shall show that such phenomenon {\em arises in the
(physical) Coulomb gauge of an abelian $U(1)$ gauge symmetry}, both in the case
of unbroken and broken $U(1)$ symmetry.

In Section 4, we show that in the case of unbroken symmetry, the (time
independent) generation of the $U(1)$ symmetry can be obtained by a suitable
time average of the integral of the charge density, through a modified Requardt
prescription~\cite{R}. This allows for a {\em direct proof of the charge
superselection rule in the (physical) Coulomb gauge} (a previous proof relied
on the general assumptions of the Feynman-Gupta-Bleuler gauge~\cite{FSW1}).

In Section 5, we discuss the implications  of the breaking of the $U(1)$ gauge
symmetry on the energy-momentum spectrum. By exploiting the
Dirac-Symanzik-Steinmann (DSS) construction of Coulomb charged
fields~\cite{DSS, BDM} we shall obtain a general non-perturbative version of
the Higgs phenomenon~\cite{WE}: {\em The (time independent) $U(1)$ gauge
symmetry is generated by the integral of the charge density discussed in
Sect.\,4, and in this case  unbroken, if and only if the Fourier transform of
the two point function of $\Fmn$ has a contribution $\d(k^2)$, i.e. there are
massless vector bosons.

If the $U(1)$ gauge symmetry is broken, then it cannot be generated (in the
above sense) by the current $j_\mu = \partial^\nu F_{\mu \nu} $; in this case
the vacuum expectation $\limR < [\,j_0(f_R, t),\,A\,] >$, where  $A$ is  a
charged field with $< A > \neq 0$, cannot vanish nor be time independent and
its Fourier spectrum coincides with the energy spectrum at $\k \ra 0$ of the
two point function of $\Fmn$; this cannot have a $\d(k^2)$ contribution, so
that the absence of massless Goldstone bosons coincides with the absence of
massless vector bosons.}

The strict analogy  of the above mechanism with the evasion of the Goldstone
theorem in  non-relativistic Coulomb systems is discussed in Sect.\,6. In
Sect.\,7 we discuss the $U(1)$ problem in QCD; we show (on the basis of local
gauges) that the {\em axial $U(1)$ transformations define a symmetry of the
observable field algebra} and argue that {\em its spontaneous breaking is not
accompanied by massless Goldstone bosons as a consequence of the time
dependence of the corresponding charge commutators in the (physical) Coulomb
gauge}.

%%%%%%%%%%%%%%%%%%%%%%%%%%%%%%%%%%%%%%%%%%%%%%%%%%%%%%%%%%%%%%%%%%%%%%%%%%%
%%%%%%%%%%%%%%%%%%%%%%%%%%%%%%%%%%%%%%%%%%%%%%%%%%%%%%%%%%%%%%%%%%%%%%%%%%%%

\section{Locality and symmetries in Quantum Field Theory} In the Wightman
formulation of Quantum Field Theory (QFT)~\cite{SW} one of the basic assumption
is that the field algebra $\F$  satisfies {\em  microscopic causality}, also
called {\em locality}; this means that fields commute or anticommute at
spacelike separations (depending on their spin). While  microscopic causality
is a must for the subalgebra $\F_{obs}$ generated by observable fields,  there
is no cogent physical reason for the locality of the whole field algebra, which
typically involves non-observable fields (like e.g.  fermion fields or charged
fields).

The locality condition for $\F$ may be read as a statement about the {\em
localization of the states} obtained by applying $\F$ to the vacuum. Following
Doplicher, Haag and Roberts (DHR)~\cite{HAAG}, a state  $\o$, defined by its
expectations $\o(A)$ on the observables, $A \in \A_{obs}$,  is {\em localized}
in the (bounded) space-time region $\O$ (typically a double cone), if for all
observable $A$ localized in the space time complement $\O'$ of $\O$, i.e. in
the set of points which are spacelike to every point of $\O$, briefly $\forall
A \in \A_{obs}(\O')$, $\o$ coincides with the vacuum state $\omega_0$ $$\o(A) =
\omega_0(A), \,\,\,\forall A \in \A_{obs}(\O').$$ Now, for any bounded region
$\O$ the unitary operators $U_{\O}$ constructed in term of fields localized in
$\O$, typically $\ph(f)$, with supp $f \subseteq \O$, give states $\o(A) \eqq
(U_\O \Psio, \, A \,U_\O \Psio)$ which are localized in $\O$ in the DHR sense.
In fact, thanks to the locality of the field algebra, $U_\O$ commutes with
$\A_{obs}(\O')$ and therefore $\o(A) = \o_0(A)$, $\forall A \in \A_{obs}(\O')$.

It is important to stress that in general the vacuum sector $\H_0$, obtained by
applying the observable field algebra $\F_{obs}$ to the vacuum, does not
exhaust the physically interesting  states and the non observable fields of
$\F$ play  the important  role of producing from the vacuum the physical states
which do not belong to $\H_0$. The properties of the non observables fields are
therefore physically interesting and worthwhile to study in view of the  states
they produce; technically the unitary operators constructed in terms of non
observables fields intertwine between the vacuum sector and the other
physically relevant representations of the observable algebra. The locality
property of $\F$ guarantees that such intertwiners are localizable and so are
the corresponding states.

In general, a one parameter group of  {\em internal  symmetries} $\b^\l, \,\l
\in \Rbf$, is a group of field transformations, technically a one parameter
group of *-automorphisms of the field algebra, which commutes with the
space-time translations $\alpha_{\x, t}$. The relation between symmetries and
localization is formalized by the following property: $\b^\l$ is {\em locally
generated } on the field algebra $\F$ if
\newline i) there exists a conserved current field $j_\mu$, $\dum j_\mu(x) =
0$,
\newline ii) the infinitesimal transformation of the field algebra is given by
\be{ \d F = i \limR [ Q_R, \, F], \,\,\,\forall F \in \F,}\ee where $Q_R$ is
morally the integral of the charge density $j_0$ in the sphere of radius $R$,
suitably regularized  to cope with the (possible) distributional UV
singularities of $j_\mu$, see eq.\,(1.2).

 The existence of a conserved current may be taken as equivalent to the
invariance of the Lagrangean (or the action); however, condition ii) is in
general not obvious, even if it is often taken for granted. Here,  locality
plays a crucial role. In fact, if the field algebra is local both the limit $R
\ra \infty$ exists {\em and } it is independent of the time smearing,
equivalently $ \limR [ j_0(f_R,t), F \,]$ is independent of time \be{\limR
\partial_t\,[ j_0(f_R,t), \,F\,] = 0,}\ee so that  eq.\,(2.1) may be checked by
(canonical) equal-time commutators.

If the  vacuum expectations of the fields, which, by the cluster property,
describe their mean behaviour at space infinity,  are invariant under $\b^\l$,
the symmetry is globally realized in the universe described by the given
vacuum, i.e. one has a global conservation law, whereas if some expectation is
not invariant, $< \d F
> \neq 0$, the symmetry is spontaneously broken and there is no global charge
associated to the current continuity equation.

For locally generated symmetries, the spontaneous  symmetry breaking implies a
strong (non-perturbative) constraint on the energy momentum spectrum, namely
the existence of massless particles, called {\em Goldstone bosons}, with the
same (conserved) quantum numbers of the current and of the field $F$ with non
invariant vacuum expectation, ({\em symmetry breaking order parameter}).

The  original proof of the theorem~\cite{GSW} applies to the case in which the
non-symmetric order parameter is given by a scalar (elementary) field $\ph$ and
exploits the  Lorentz covariance of the two point function $< j_\mu(x) \ph(y)
>$, but it was later realized that the crucial property is the {\em relative
locality} between $j_\mu$ and the symmetry breaking  order parameter $F$,
(which needs not to be one of the basic or elementary fields, but  may be a
polynomial of them)~\cite{KRS}. The lack of appreciation of this point has been
at the basis of discussions and  attempts for evading the Goldstone theorem,
which eventually led to the Higgs mechanisms and to the standard model of
elementary particles.

%%%%%%%%%%%%%%%%%%%%%%%%%%%%%%%%%%%%%%%%%%%%%%%%%%%%%%%%%%%%%%%%%%%%%
%%%%%%%%%%%%%%%%%%%%%%%%%%%%%%%%%%%%%%%%%%%%%%%%%%%%%%%%%%%%%%%%%%%%%%
%%%%%%%%%%%%%%%%%%%%%%%%%%%%%%%%%%%%%%%%%%%%%%%%%%%%%%%%%%%%%%%%%%%%%
\section{Locality and symmetries in gauge theories}
 Gauge field theories  exhibit very distinctive features, with fundamental
 experimental consequences,
like spontaneous symmetry breaking with energy gap (Higgs mechanism) in
apparent contradiction with Goldstone theorem, quark confinement and linearly
rising potential in contrast with the cluster property, axial current anomaly,
asymptotic freedom etc. \footnote{For the general structure and properties of
gauge field theories see~\cite{W}. For the lack of locality and the violation
of cluster property see~\cite{FS0}; for a non-perturbative discussion of the
evasion of the Goldstone theorem in gauge theories see~\cite{ST,FSGM}.}

It is natural to try to understand such departures from standard quantum field
theory in terms of general ideas independently  of the specific model. The
original motivation by Yang and Mills, namely  that quantum numbers or charges
associated to gauge transformations have only a local meaning  does not have a
direct experimental interpretation since,  as a consequence of confinement and
symmetry breaking,  the observed physical states do not carry non abelian gauge
charges. More generally, by definition gauge transformations reduce to the
identity on the observables, so that they can be defined only by introducing
non observable fields. The role of gauge symmetries has therefore been
regarded~\cite{HAAG, D} as that of providing a classification of the
(inequivalent) representations of the observable algebra, through the action of
the charged fields. It is still  unexplained why only states corresponding to
one dimensional representations of the gauge groups (which include the non
abelian gauge group of permutations of identical particles) occur in nature.

For the meaning of local gauge invariance, we recall that the standard
characterization of gauge field theories is that they are formulated in terms
of (non observable) fields which transform non trivially under the group $\G$
of {\em local gauge transformations} leaving the Lagrangean invariant.

In classical field theory, the invariance of the Lagrangean or of the
Hamiltonian under  a ($n$-dimensional) Lie group $G$ of space time independent
field transformations can be checked by considering the infinitesimal variation
of the fields $\ph_i$ (for simplicity we  take $G$ compact and include the
coupling constants in the generators)
\def \da {\d^{(a)}}
\def \de {\d^\eps}
 \be{\d \ph_i(x) = i \eps_a t^a_{i\, j}\, \ph_j(x) \eqq i (\eps
t\,\ph)_i(x),}\ee \be{\d A^a_\nu(x) = i \eps_c T^c_{a \, b}  A_\nu^b(x) \eqq i
(\eps T A_\nu)^a(x),\,\,\, T^a_{b\, c} =
 i f^b_{a\,c}}\ee where $a = 1, ...n$, $i = 1, ...d $,  summation over
repeated indices is understood, $\eps$ are the infinitesimal group parameters,
$t$ is the ($d$-dimensional) matrix representation of the generators of the
group $G$, provided by the fields $\ph_i$ and $f$ are the Lie algebra structure
constants.

The {\em local gauge group $\G$ associated to $G$} ( called the {\em  global
group}), is the infinite dimensional group obtained by letting the group
parameters to be regular {\em localized} functions $\eps(x)$ of the space time
points, typically $\eps \in \D(\Rbf^4)$ or $\in \S(\Rbf^4)$, with the result of
an additional term $\dun \eps^a(x)$ in eq.\,(3.2).

It is very important to keep separate the Lie algebra $L(G)$ of $G$
 and the infinite
dimensional algebra corresponding to $\G$, briefly denoted by $L(\G)$. It would
be improper to consider the first as a finite dimensional
 subalgebra of the
second, both from a mathematical and for a physical point of view. In
particular, trivial representations of the $L(\G)$ need not to be trivial
representations of $L(G)$ and in fact the construction of gauge invariant
charged fields is one of the strategies for the analysis of gauge theories; an
example of such a construction is the DSS construction in the abelian case of
the DSS fields~\cite{DSS}.

\def \dlf {\frac{\d \L}{\d \ph_i}}
\def \dldf {\frac{\d \L}{\d \dmu \ph_i}}
\def \dla  {\frac{\d \L}{\d A_\nu^b}}
\def \dlda {\frac{\d \L}{\d \dmu A_\nu^b}}
A physically very important consequence of the invariance under a local gauge
group is that one gets a stronger form of the local conservation laws $\dum
J^a_\mu(x) = 0$, implied by  the invariance under the  {\em global group} $G$.
In fact, by the second Noether theorem, the conserved currents

  \be{J^{a\,\mu}
\eqq - i \frac{\d \L}{\d \dmu \ph_i} (t^a \ph)_i - i \dlda (T^a A_\nu)^b \eqq
j^{a\,\mu}(\ph) + j^{a \,\mu}(A),}\ee satisfy the additional equation \be{
J^b_\mu = \frac{\d \L}{\d A^\mu_b} =  \partial^\nu\,
 G^b_{\mu \,\nu} + E[A]^b,\,\,\,\,\,\,G^{\mu\,\nu}_b \eqq - \dlda = - G^{\nu\,
\mu}_b,}\ee where $E[A]_b = 0$ are  the Euler-Lagrange (EL) equations of motion
of  $A$.

 The equations (3.4) encode the invariance under the local gauge group
$\G$, briefly  {\em gauge invariance}, and can be taken as a characterization
of such an invariance property; they shall be called {\em local Gauss' laws},
since Gauss' theorem represents their integrated form. The current continuity
equation is trivially implied without using the EL equations for the matter
fields.

The validity of local  Gauss' laws appears to have a more direct physical
meaning than the gauge symmetry, which is non trivial  only on non observable
fields. It is therefore tempting to regard the validity of local Gauss' laws
 as the basic characteristic feature of gauge field theories, and to  consider
gauge invariance merely as a useful recipe  for writing down Lagrangean
functions which automatically lead to the validity of  local Gauss' laws.

Actually, for the canonical formulation of gauge field theories one has to
exploit the freedom of fixing a gauge, typically  by adding a gauge fixing term
in the Lagrangean (irrelevant for the physical implications), and this can be
done even at the expense of totally breaking the gauge invariance of the
Lagrangean, (as e.g. in the so called unitary gauge). Thus, the gauge
invariance of the Lagrangean is not so crucial from a physical point  of view,
whereas so is the validity of local Gauss' laws, which is preserved under the
addition of a gauge fixing and the corresponding subsidiary condition~\cite{W}.

As we shall see below, the local Gauss' law is at the basis of most of the
peculiar features of gauge quantum field theories, with respect to standard
quantum field theories.\footnote{The recognition of
 local Gauss' laws as the basic characteristic features
of gauge field theories has been argued and stressed (also in view of
 the quantum theories) in~\cite{FSW} and later re-proposed~\cite{KK},
ignoring the above references.}

 From a structural point of view, a first
consequence of the local Gauss' law is that, if the local charges $Q^a_R$,
eq.\,(1.2), generate the global group $G$,  the {\em charged fields cannot be
local}~\cite{FPS}. In fact, the field $F$ is charged with respect to the $a$-th
one parameter subgroup of $G$ if $\d^a F \neq 0$, whereas if $F$ is {\em local}
with respect to the conserved current $J^a_\mu$, \be{\limR [ Q^a_R, F ] = \int
d^3x d t \,\nabla_i
 f_R(\x) \a(t) [\,G_{0 \,i}(\x,t), F\,] =0,}\ee since  supp\,$\nabla_i
 f_R \a \subset \{ R \leq |\x| \leq R(1 + \eps) \}$
 becomes spacelike with respect to any bounded region for $R$ large
 enough. The r.h.s. vanishes also for more general time
 smearing, $\a_R(t)$ with support in $[-R(1-\eps), \,R(1-\eps) ]$ (see below).

 Since the local generation of $G$ follows from the locality of the time
 evolution and of the equal time canonical commutators, eq.\,(3.5) implies
 that the charged fields are not local with respect to $G_{i\,0}$.

The physical reason is that  Gauss' law establishes a tight link between the
local properties of the solutions and their behaviour  at infinity; e.g. the
charge of a solution of the electrodynamics equation can be computed either by
integrating the charge density, i.e. a local function of the charge carrying
fields, or by computing the flux of the electric field at space infinity.

 This result has very strong implications at the level of structural
properties of  gauge quantum field theories:  {\em the  field algebra}
generated by $G^a_{\mu\,\nu}$ and the charged fields {\em cannot be local}.

This may appear as a mere gauge artifact with no physical relevance,~\cite{H}
since charged fields are not observable fields. However, charged fields play
the important role of generating from the vacuum  charged states and describing
(even neutral) states in terms of charged particles. The non locality of the
charged fields, as implied by the local Gauss' law, has therefore the important
physical consequence that the charged states cannot be local in the DHR sense.

%%%%%%%%%%%%%%%%%%%%%%%%%%%%%%%%%%%%%%%%%%%%%%%%%%%%%%%%%%%%%%%%%%%%%%%%
%%%%%%%%%%%%%%%%%%%%%%%%%%%%%%%%%%%%%%%%%%%%%%%%%%%%%%%%%%%%%%%%%%%%%%%
%%%%%%%%%%%%%%%%%%%%%%%%%%%%%%%%%%%%%%%%%%%%%%%%%%%%%%%%%%%%%%%%%%%%%%%%%%%

\section{Gauss' law and local generation of symmetries} Since the fields which
transform non trivially under  the global group $G$ cannot be local, the local
generation  of $G$ becomes problematic, namely both the existence of the limit
$R \ra \infty$ in eq.\,(1.2), as well as  its time independence are in question
and, as far as we know, no general conclusion follows directly from Gauss'
law. As we shall show, both questions can be answered for Coulomb charged
fields, by (crucially) exploiting their construction in terms of the local
charged fields of the Feynman-Gupta-Bleuler (FGB) gauge.

\begin{Proposition} In the Coulomb gauge of QED, $\forall \Psi, \,\Phi \in \F_C
\Psio$, with $\F_C$ the field algebra of the Coulomb gauge and $\Psio$ the
vacuum vector, the limits \be{ \limR (\Psi, \,[\,j_0(f_R\,\a),
\,F\,]\,\Phi)}\ee exist; however, they are (generically) $\a$ dependent if $F$
is a charged field and therefore eq.\,(1.2) fails.
\end{Proposition}
\Pf  The field algebra $\F_C$  is generated by the vector potential $A_C^i$ and
the elementary charged fields $\ph_C$, so that  it is enough to discuss the
case $F = \ph_C$ and a basic ingredient is the DSS construction~\cite{DSS} of
the Coulomb charged fields $\ph_C$ in terms of the local  fields $\ph, \,A_\mu$
of the FGB gauge
  \be{\ph_C(y) = e^{ie (-\Delta^{-1} \partial^j A_j)(y)}\,\ph(y).}\ee
The necessary ultraviolet regularization of eq.\,(4.2) has been discussed by
Steinmann~\cite{DSS} within the perturbative expansion. A regularized version
which only uses the existence of the FGB correlations and it is constructed by
using in the exponential fields smeared in space an time, has been given
Buchholz et al.~\cite{BDM}. In this framework, the space asymptotic of the
correlation functions of the commutator  $[\, \Fmn (x),\, \ph_C(y)\,]$ is
given, at all orders in the expansion of the exponential entering in
eq.\,(4.2), with corrections $ O(|\x|^{-4})$, by \be{ [\, \Fmn(x), \ph_C(y)]
\sim \frac{-ie}{4\pi} \int d^3 z \,\partial_z^j \frac{1}{|\z - \y |} <[\,
\Fmn(x),\, A_j(\z, y_0)]> \,\ph_C(y). }\ee
\def \do {{\partial_0}}
Since $<[\, \Fmn(x),\, A_j(z)]> = i( \dnu g_{\mu\,j} - \dmu g_{\nu \,j})
K(x-z)$, with $K$ the commutator function of the electromagnetic field, one
has, for $R \ra \infty$, \be{[j_0(f_R,x_0), \ph_C(y)] = [\partial^i F_{0
i}(f_R,x_0), \, \ph_C(y)] \sim - e \do \int d^3 x\,f_R(\x) K(x-y) \ph_C(y).}\ee
 By the support properties of $K(x) = -i \int d \rho(m^2) \,\eps(k_0) \,\d(k^2
- m^2)\,e^{-i k x}$, the charge density is integrable and, in all correlation
functions, \be{ \limR [\,j_0(f_R, x_0),\, \ph_C(y)\,] = e \int d \rho(m^2)\,
\cos (m (x_0- y_0))\,\ph_C(y).}\ee The r.h.s is independent of time if and only
if $d \rho(m^2) = \l \d(m^2)$, i.e. if $\Fmn$ is a free field.

The same conclusions are obtained if instead of eq.\,(4.2) one uses the
regularized version of Ref.~\cite{BDM}, since in this case eqs.\,(4.3-5) get
changed only by a convolution with a test function $h(\y_0) \in \D(\Rbf)$.

 \vspace{1mm} The time dependence of $\limR [j_0(f_R, t),
F]$, $ F \in \F_C$ is compatible with the conservation of the current because
the above analysis gives $$[j_i(x),\,\ph_C(y)\,] =  (e/4 \pi) \int d^3 z
\partial_z^i |\z - \y|^{-1} \do^2 K(x-y),$$  \be{\limR [\,\dot{Q}_R(x_0),
\ph_C(y)\,] = [\, \mbox{div} \j(f_R, x_0),\,\ph_C(y)\,] \neq 0.}\ee  The time
dependence of the commutator of eq.\,(4.5) is at the basis of the appearance of
an infinite renormalization constant in the equal time commutator of the charge
density $j_0 = \partial^i F_{0\, i}$ and the Coulomb charged field $\ph_C$
$$[\,j_0(x), \,\ph_C(y)\,]_{x_0 = y_0} =  e (Z_3)^{-1}\, \d(\x-\y)
\,\ph_C(y),$$ (all fields being renormalized fields and $e$ the renormalized
charge), as it appears by comparing the integrated form of the above equal time
commutator and eq.\,(4.4). For such a phenomenon the vacuum polarization due to
fermionic loops plays a crucial role, so that the semi-classical approximation
does not provide relevant information and in fact the phenomenon does not
appear in the classical theory.

Proposition 4.1 shows that, in contrast with the local case, the equal time
commutators are misleading for the charge commutators and, contrary to
statements in the literature, {\em the $U(1)$ charge group of QED is not
locally generated by the integral of the charge density, in the sense of
eq.\,(4.1)}. Thus,  the heuristic argument that if the symmetry commutes with
the time translations, equivalently if the current continuity equation holds,
then the generating charge commutes with the Hamiltonian  and is therefore
independent of time is not correct. Time independence of the charge commutator
holds provided one has (relative) locality  at all times between $j_\mu$ and
the charged fields; now, even if the equal time commutators have a sufficient
localization, the time evolution may induce a delocalization leading to a
failure of eq.\,(2.2).
%%%%%%%%%%%%%%%%%%%%%%%%%%%%%%%%%%%%%%%%%%%%%%%%
%%%%%%%%%%%%%%%%%%%%%%%%%%%%%%%%%%%%%%%%%%%%%%%%%%%%%%%%%%%%%
%%%%%%%%%%%%%%%%%%%%%%%%%%%%%%%%%%%%%%%%%%%%%%%%%%%%%%%%%%%%%%%%
%%%%%%%%%%%%%%%%%%%%%%%%%%%%%%%%%%%%%%%%%%%%%%%%%%%%%%%%%%%%%
\section{Electric charge and its superselection}
The results of the previous section leave open the question of whether a
modification of eq.\,(1.2) may yield a relation between a gauge symmetry and
the charge density of the corresponding Noether current. As we shall discuss
below, if the gauge symmetry is unbroken a time average of eq.\,(1.2), similar
to that proposed by Requardt~\cite{R, B}, provides the required relation.

Actually,  eq.\,(4.4) gives the renormalized charge for any time smearing
$\a_{T(R)}(x_0) \eqq \a(x_0/T(R))/T(R)$, with $T/R \ra 0$ as $R \ra
\infty$~\cite{MS}, if $d \rho(k^2)$ has a $\d(k^2)$ contribution. Moreover, for
a certain class of functions $T(R)$, which depends on the infrared behaviour of
$k^2\,d \rho(k^2)$,  $j_0(f_R\, \a_{T(R)}) \Psio $ converges strongly to
zero~\cite{MS}. A smearing which gives both results independently of any
information on the above infrared behaviour, is given by taking $ T(R) = \d R$,
with $ \d \ra 0$ after the limit $R \ra \infty$.

\begin{Proposition} In the Coulomb gauge the $U(1)$ gauge symmetry is generated
by the integral of the charge density  \be{ \d F = i \lim_{\d \ra 0} \limR
[\,Q_{R\, \d}, \,F\,],}\ee \be{Q_{R\,\d} \eqq j_0(f_R \,\a_{\d R}),
\,\,\,\,\,\a_{\d R}(x_0) \eqq \a(x_0/(\d R))/(\d R)}\ee if and only if $d
\rho(k^2)$ has a $\d(k^2)$ contribution, i.e. there are massless photons.

Moreover, one has \be{ \mbox{strong}- \limR j_0(f_R\,\a_{\d R})\,\Psio = 0,}\ee
so that, it there are massless photons one can express the electric charge $Q$,
i.e. the generator of the $U(1)$  symmetry, as an integral of the charge
density $j_0$ not only in the commutators with charged fields, but also in the
matrix elements of the Coulomb charged states $\Phi, \,\Psi \in \F_C \Psio$
\be{ (\Phi, Q\,\Psi) = \lim_{\d \ra 0}\limR (\Phi, \,j_0(f_R\,\a_{\d
R})\,\Psi).}\ee
\end {Proposition}
\Pf  The time  smearing  of eq.\,(4.4) with $\a_{\d R}(x_0)$ gives
 $$ [\,j_0(f_R \a_{\d R}),\, \ph_C(y)\,] =$$ $$=
 e \int d \rho(m^2)\,d^3 q \,\tilde{f}(\q) \, \mbox{Re}[e^{-i
\o_R(q, m)y_0}\tilde{\a}(\d \sqrt{\q^2 + R^2 m^2})]\, \ph_C(y),$$ where
$\o_R(\q,m)\eqq \sqrt{\q^2 R^{-2}+m^2}$. Then, since $\a$ is of fast decrease,
by the dominated convergence theorem the r.h.s. vanishes if the $d \rho(m^2)$
measure of the point $m^2 = 0$ is zero, i.e. if there is no $\d(m^2)$
contribution to $d \rho$. In general, if the point $m^2$ has measure $\l$,  one
gets $ \l e \,\ph_C(y)$; finally the renormalization condition of the
asymptotic electromagnetic field gives $\l = 1$.

For the proof of eq.\,(5.3) one has ($d \Omega_m(\k) \eqq d^3k ( 2 \sqrt{\k^2 +
m^2})^{-1}$) $$||Q_{R \d}\,\Psio||^2 = \int d \rho(m^2) m^2 d \Omega_m(\k)\,
|\, \k\, \tilde{f}_R(\k) \tilde{\a}(\d R \sqrt{\k^2 + m^2})|^2 =$$ $$= \int d
\rho(m^2)\,d \Omega_m(\q/R)\, m^2\,R \, |\tilde{\a}(\d \sqrt{\q^2 + m^2
R^2})\,\q\, \tilde{f}(\q)|^2.$$ Now,  $m^2 R \,|\tilde{\a}(\d \sqrt{\q^2 + m^2
R^2})|^2$ converges pointwise to zero for $R \ra \infty$ and since $d \rho(m^2)
$ is tempered and $\a$ is of fast decrease the r.h.s. of the above equation
converges to zero by the dominated convergence theorem.

 \vspace{2mm}One of the basic Dirac-Von
Neumann axioms of quantum mechanics is that the states of a quantum mechanical
system are described by vectors of a Hilbert space $\H$ and that every vector
describes a state, equivalently all projections and therefore all (bounded)
self-adjoint operators  represent observables (briefly $\A_{obs} = \B(\H)$). It
was later realized~\cite{WWW} that, typically for systems with infinite degrees
of freedom,  the physical states may belong to a direct sum of irreducible
representations of the observable algebra and therefore one cannot measure
coherent superpositions of vectors belonging  to inequivalent representations
of the observable algebra. This means that if $\H = \oplus \H_j$, each $\H_j$
carrying an irreducible representation of $\A_{obs}$, a linear combination $\a
\Psi_1 + \b \Psi_2$ of vectors $\Psi_1, \,\Psi_2$ belonging to different $\H_j$
is not a physically realizable (pure) state and it rather describes a mixture
with the density matrix $|\a|^2 \Psi_1 \otimes \Psi_1 + |\b|^2 \Psi_2 \otimes
\Psi_2$.

The impossibility of measuring the relative phase of a linear combination of
vectors is equivalent to the existence of  operators $Q$, called {\em
superselected charges}, which commute with all the observables (and have a
denumerable  spectrum if the Hilbert space is separable). \footnote{The
superselected charges are often called {\em gauge charges}, but we prefer the
name of superselected charges.
 The gauge group which classify the representations  of the observable algebra
defined by DHR localized states has been proved to be compact~\cite{D}.}

Wick, Wightman and Wigner (WWW) proved that rotation and time reversal
invariance imply that the operator $Q_F = (-1)^{2 J} = (-1)^F$ where $J$ is the
angular momentum and $F$ is the fermion number modulo $2$ is a superselected
charge ({\em univalence superselection rule}, also called {\em fermion-boson
superselection rule}). It was later shown that only rotational invariance was
needed for the proof~\cite{HKW}. WWW also suggested that the electric charge
and possibly the baryon number define superselected charges.

The superselection rule for the electric charge was later questioned and
debated~\cite{MAS}. The proof may be dismissed as trivial by arguing that
observables must be gauge invariant and that gauge invariance imply zero
charge, but as stressed before such an argument is not correct, since the
latter implication is contradicted by the Dirac-Symanzik-Steinmann field
operator~\cite{DSS} showing that gauge invariant operators need not to commute
with the electric charge.

The superselection of the electric charge $Q$ may be shown to be a consequence
of the locality of the observables and  the Gauss law, provided one can express
$Q$ as an integral of $j_0 = \partial^i F_{0 i}$. A proof of the charge
superselection rule has been given  by using a local gauge quantization of QED,
like e.g. the Feynman-Gupta-Bleuler gauge, and by identifying $Q$ with the
generator of the global gauge transformations of the local fields~\cite{FSW}.
In this gauge, the  construction of the DSS operators~\cite{MS}  makes clear
that {\em invariance under the local gauge transformations does not imply
invariance under the global gauge transformations for non local operators}.

By exploiting Prop.\,5.1 one can get a direct proof of the charge
superselection rule in the {\em physical} Coulomb gauge.
\begin{Proposition} The electric charge $Q$, defined in the Coulomb gauge by $$
Q \,\Psio = 0, \,\,\,\,\,[\, Q,\,\ph_C(y)\,] = e \,\ph_C(y),$$ commutes with
the observables (on the Coulomb states) \be{ ( \Phi, \, [ Q,\, A ] \Psi ) =
\lim_{\d \ra 0}\,\limR ( \Phi, \,[ j_0(f_R\,\a_{\d R}), \, A ] \Psi ) = 0,
\,\,\,\,\,\forall \Phi, \,\Psi \in \F_C\, \Psio,}\ee and it is therefore
superselected. \end{Proposition} \Pf The proof follows from eq.\,(5.4), which
relates the electric charge $Q$ and the electric flux at infinity,  by the
argument which exploits  the relative locality of the observables with respect
to the (observable) electromagnetic field, as required by Einstein
causality,~\cite{FSW1, B} so that the r.h.s. vanishes by the same argument of
eq.\,(3.5). Actually the r.h.s. of eq.\,(5.5) vanishes independently of the
adopted time smearing by locality.

The superselection of electromagnetic fluxes at spacelike infinity has been
discussed by Buchholz~\cite{B}, under the assumption of weak convergence. The
special choice of space time smearing $j_0(f_R \a_{\d\,R})$ adopted above
guarantees the strong convergence on the vacuum, convergence in expectations on
Coulomb charged states
 and the relation between the corresponding electric flux
and the electric charge.
%%%%%%%%%%%%%%%%%%%%%%%%%%%%%%%%%%%%%%%%%%%%%%%%%%%%%%%
%%%%%%%%%%%%%%%%%%%%%%%%%%%%%%%%%%%%%%%%%%%%%%%%%%%%%%%%%%%%%%
%%%%%%%%%%%%%%%%%%%%%%%%%%%%%%%%%%%%%%%%%%%%%%%%%%%%%%%%%%%%%%
%%%%%%%%%%%%%%%%%%%%%%%%%%%%%%%%%%%%%%%%%%%%%%%%%%%%%%%%%%%%%%%%

\section{Gauge symmetry breaking and the energy-momentum spectrum.
The Higgs mechanism} The Higgs mechanism,  relative to the breaking of the
global group $G $ in a gauge quantum field theory, plays a crucial role in the
standard model of elementary particle physics. The standard discussion of this
mechanism is based on the perturbative expansion and, in particular, the
evasion of the Goldstone theorem is  checked at the tree level with the
disappearance of the massless Goldstone bosons and the vector bosons becoming
massive~\cite{HGS}. This is displayed by the Higgs-Kibble (abelian) model
\index{Higgs-Kibble model} of a (complex) scalar field $\ph$ interacting with a
real gauge field $A_\mu$, defined by the following Lagrangean ($\rho(x) \eqq
|\ph(x)|$) \be{\L = - {{\scriptstyle{\frac{1}{4}}}} \Fmn^2 + \ume |D_\mu \ph|^2
- U(\rho),
 \,\,\,\,D_\mu = \dmu -i e A_\mu}\ee  invariant under the $U(1)$ gauge
group: $\b^\l(\ph) = e^{i \l }\ph$, $\b^\l(A_\mu) = A_\mu$ and under local
gauge transformations.

At the classical level, one may argue that by a  local gauge transformation
$$\ph(x) = e^{i \theta(x)}\,\rho(x) \ra \rho(x),\,\, A_\mu(x) \ra A_\mu(x) +
e^{-1} \dmu \theta(x) \eqq W_\mu(x)$$ one may eliminate the field $\theta$ from
the Lagrangean, which becomes \be{\L =  - {{\scriptstyle{\frac{1}{4}}}} \Fmn^2
+ \ume e^2 \rho^2\,W_\mu^2 + \ume (\dmu \rho)^2 - U(\rho).}\ee

\def \bph {{\overline{\ph}}}
\def \bro {{\overline{\rho}}}
If the (classical) potential $U$ has a non trivial (absolute) minimum $\rho =
\bro$ one can consider a semiclassical approximation based on the expansion
$\rho = \bro + \s$, treating  $\bro$ as a classical constant field and $\s$ as
small. At the lowest order, keeping only the quadratic terms in $\s$ and
$W_\mu$
  one has  \be{\L^{(2)} =
 - {{\scriptstyle{\frac{1}{4}}}} \Fmn^2 + \ume e^2
\bro^2\,W_\mu^2 + \ume (\dmu \s)^2 - \ume U''(\bro) \s^2.}\ee This  Lagrangean
describes a massive vector boson and a massive scalar with (square) masses
$M_W^2 = \ume e^2\,\bro^2$, $m_\s^2 = U''(\bro)$, respectively. This argument
is commonly taken as an evidence that there are no massless particles in the
theory described by the Lagrangean $\L$.

This argument, widely used in the literature, is not without problems, because
already at the classical level, for  the equivalence between  the two forms  of
the Lagrangean, eqs.\,(6.1),(6.2)), one must add the constraint that $\rho$ is
positive, a property which is in general spoiled by  the time evolution given
by the Lagrange equations for the variables $\rho$ and $W_\mu$. For the
variables of the quadratic Lagrangean (6.3), one should require that the time
evolution of $\s$ keeps it bounded by $\bro$, a condition which is difficult to
satisfy. Thus, the constrained system is rather singular and its mathematical
control is doubtful. The situation becomes obviously more critical for the
quantum version, since the definition of $|\ph(x)|$ is very problematic also
for distributional reasons. In conclusion,  $\rho$ is a very singular field and
one cannot consider it as a genuine Lagrangean (field) variable.

A better alternative is to decompose the field $\ph = \ph_1 + i\,\ph_2$ in
terms of hermitian fields, and to consider the semiclassical expansion $\ph_1 =
\bph + \chi_1$, $\ph_2 = \chi_2$, treating $\chi_i, \,i = 1, 2$, as small. By
introducing  the field $W_\mu \eqq A_\mu +e^{-1} \dmu \chi_2$,
 one eliminates  $\chi_2$  from the quadratic part of the so expanded
Lagrangean, which gets exactly the same form  of eq.\,(6.2), with $\bro$
replaced by $ \bph$ and $\s$ by $\chi_1$.

If indeed the fields  $\chi_i$ can be treated as small, by appealing to the
perturbative (loop) expansion one has that $<
 \ph > \sim \bro \neq 0$, i.e. the vacuum
expectation of $\ph$ is not invariant the $U(1)$ charge group ({\em symmetry
breaking}). Thus, the expansion can be seen as an expansion around a (symmetry
breaking) mean field ansatz and it is very important that a renormalized
perturbation theory based on it exists and yields a non vanishing symmetry
breaking order parameter $< \ph > \neq 0 $ at all orders. This is the standard
(perturbative) analysis of the Higgs mechanism.

 The extraordinary success of the standard model motivates an examination of
the Higgs mechanism from a general non-perturbative point of view. In this
perspective, one of the problems is that mean field expansions may yield
misleading results about the occurrence of symmetry breaking and the energy
spectrum. \footnote{E.g. the mean field ansatz on the Heisenberg  spin model of
ferromagnetism gives a  wrong critical temperature and an energy gap. For a
discussion of the problems of the mean field expansion see e.g.~\cite{FSSB}.}
Actually, a non-perturbative analysis of  the euclidean functional integral
defined by the Lagrangean of eq.\,(6.1),
 gives symmetric correlation functions and in
particular $< \ph > = 0$ ({\em Elithur-De Angelis-De Falco-Guerra (EDDG)
theorem}~\cite{EG}). This means that the mean field ansatz  is incompatible
with quantum effects and the approximation leading to eq.\,(6.3) is not
correct.\footnote{The crux of the argument is that gauge invariance decouples
the transformations of the fields inside a volume $V$ (in a euclidean
functional integral approach) from the transformation of the boundary, so that
the boundary conditions are ineffective and cannot trigger non symmetric
correlation functions. For a simple account of the argument see
e.g.~\cite{FS}.}

The same negative conclusion would be reached if, (as an alternative to the
transformation which leads to eq.\,(6.2)), by means of a gauge transformation
one reduces $\ph(x)$ to a real, not necessarily positive, field $\ph_r(x)$.
This means that the local gauge invariance has not been completely eliminated
and the corresponding Lagrangean, of the same form (6.2) with $ \rho$ replaced
by $\ph_r$, is invariant under a residual $Z_2$ local gauge group. An easy
adaptation of the proof of the EDDG theorem gives $ < \ph_r > = 0$.

In order to avoid the vanishing of a symmetry breaking order parameter one must
reconsider the problem by adding to the Lagrangean (6.1) a gauge fixing
$\L_{GF}$ which breaks local gauge invariance. Thus, {\em the  discussion of
the Higgs mechanism is necessarily  gauge fixing dependent}; this should not
appear strange, since the vacuum expectation of $\ph$ is a gauge dependent
quantity.

The important physical properties at the basis of the Higgs mechanism are
particularly clear in the so called physical gauges, like the Coulomb gauge.
Since the charged Coulomb fields cannot be local, the local generation of the
symmetry, required for the applicability of the Goldstone theorem, is in
question. Actually,  in the abelian case by using the results of Sect.\,5 one
has  a non perturbative proof of the characterization of the Higgs phenomenon
given by Weinberg on the basis of the perturbative expansion~\cite{W}. By
Prop.\,5.1, the (time independent) $U(1)$ gauge symmetry is generated by the
integral of the charge density, eq.\,(5.1), and in this case unbroken, if and
only if the Fourier transform of the two point function of $\Fmn$ has a
contribution $\d(k^2)$, i.e. there are massless vector bosons.
\begin{Proposition}
If the (time independent) $U(1)$ gauge symmetry is broken, then it cannot be
generated by the integral of the charge density, eq.\,(5.1), of the associated
Noether current $j_\mu = \partial^\nu F_{\nu \mu}$; in this case the vacuum
expectation $\limR < [\,j_0(f_R, t),\,A\,] >$, where  $A$ is a charged field
with $< A > \neq 0$, cannot vanish nor be time independent and its Fourier
spectrum coincides with the energy spectrum at $\k \ra 0$ of the two point
function of the vector boson field $\Fmn$, which cannot have a $\d(k^2)$
contribution, so that the absence of  massless Goldstone bosons coincides with
the absence of massless vector bosons.
\end{Proposition}
\Pf The first part  follows trivially  from Proposition 5.1, which also states
that the symmetry generated according to eq.\,(5.1) cannot be broken. Thus, in
the broken case the spectral measure of $\Fmn$ cannot have a $\d(k^2)$
contribution and therefore $\lim_{\d \ra 0} \limR [ j_0(f_R \a_{\d R}), F\,] =
0$, $ \forall F \in \F_C$. The vacuum expectation $\limR < [\,j_0(f_R,
t),\,A\,]
>$, where $ A$ is a charged field with $< A > \neq 0$, is obtained from
eqs.\,(4.4), (4.5) and the relation with the Fourier spectrum of $\Fmn$
follows.

In conclusion, the above discussion shows that the evasion of the Goldstone
theorem crucially depends on the non locality of the charged fields. The local
structure of the canonical commutation relations, in particular  of the
commutator $[ \j, \, \ph_C ]$, is not stable under  the time evolution induced
by the electromagnetic interactions, as displayed by the Coulomb gauge. This is
possible in a relativistic theory because in this case the field algebra does
not satisfy manifest covariance. For these reasons, no reliable information can
be inferred from the equal time commutators and the check of the basic
assumptions of the Goldstone theorem becomes interlaced with the dynamical
problem, as it happens for non-relativistic systems. The failure of locality
leading to eq.\,(4.6), rather than the lack of manifest covariance, is the
crucial structural property which explains the evasion of the Goldstone theorem
in the Higgs mechanism as well as in Coulomb systems and in the $U(1)$ problem,
as  discussed below.

%%%%%%%%%%%%%%%%%%%%%%%%%%%%%%%%%%%%%%%%%%%%%%%%%%%%%%%%%%%%%%%%%%%%%%%%%%%%
%%%%%%%%%%%%%%%%%%%%%%%%%%%%%%%%%%%%%%%%%%%%%%%%%%%%%%%%%%%%%%%%%%%%%%%%%%
%%%%%%%%%%%%%%%%%%%%%%%%%%%%%%%%%%%%%%%%%%%%%%%%%%%%%%%%%%%%%%%%%%%%%%%%%%
\sloppy
\section{Coulomb delocalization and symmetries in many body theory}
\fussy A natural question, following by the above discussion of symmetry
breaking, is the general characterization of the dynamics which induces a
delocalization leading to the failure of eq.\,(2.2), so that the symmetry is
not locally generated and one may have symmetry breaking with energy gap. In
this perspective, whenever the field algebra is not manifestly covariant, {\em
instantaneous} interactions are possible and there is no longer a deep
distinction between relativistic and non-relativistic systems. Actually, both
the Coulomb gauge in QED and the non-relativistic Coulomb systems are
characterized by the instantaneous Coulomb interaction \be{H_{int} =  \ume e^2
\int d^3 x\,d^3 y\, j_0(\x) V(\x - \y) j_0(\y).}\ee As argued by
Swieca~\cite{SWIE}, for two body instantaneous interactions the range of the
potential characterizes the delocalization induced by the dynamics: if $V(\x)$
falls off like $|\x|^{-d}$, for $|\x| \ra \infty$, then the unequal time
commutators generically  decay with the same power and \be{\lim_{\x \ra \infty}
|\x|^{d + \eps}[ A_{\x}, B_{t} ] = 0,}\ee at least order by order in a
perturbative expansion in time.

The above equation suggests that the  delocalization needed for eq.\,(3.9) in
three space dimensions is that given by a potential fall off like $|\x|^{ -2}$.
Actually,  the current $\j$ involves space derivatives and the critical decay
turns out to be $|\x|^{-1}$, i.e. that of the Coulomb potential.

This unifies the mechanism of symmetry breaking with energy gap of the Higgs
phenomenon and that of  non-relativistic Coulomb systems (typically the
breaking of the Galilei symmetry in the jellium model or the breaking of the
electron $U(1)$ symmetry in the BCS model of superconductivity) and provides a
clarification of the analogies proposed by Anderson~\cite{And} For a general
discussion of the energy gap associated to symmetry breaking for long range
dynamics see~\cite{FSGM}.

In gauge theories relative locality may fail because either the order parameter
(as in the Higgs phenomenon) or the conserved current (as discussed below)
associated to the symmetry of the Lagrangean {\em non-local} fields. This is
the case of the $U(1)$ problem in QCD.

%The latter is the case of the axial $U(1)$ symmetry in quantum chromodynamics).

%%%%%%%%%%%%%%%%%%%%%%%%%%%%%%%%%%%%%%%%%%%%%%%%%%%%%%%%%%%%
%%%%%%%%%%%%%%%%%%%%%%%%%%%%%%%%%%%%%%%%%%%%%%%%%%%%%%%%%%%%%%%
%%%%%%%%%%%%%%%%%%%%%%%%%%%%%%%%%%%%%%%%%%%%%%%%%%%%%%%%
%%%%%%%%%%%%%%%%%%%%%%%%%%%%%%%%%%%%%%%%%%%%%%%

\section{Axial symmetry breaking and U(1) problem}
The debated problem of $U(1)$ axial symmetry breaking in quantum chromodynamics
without  massless Goldstone bosons can be clarified by the realization of the
non locality of the associated axial current. As clearly shown by
Bardeen~\cite{BA},  the $U(1)$ axial symmetry gives rise to a conserved, gauge
dependent,  current $$J_\mu^5 = j_\mu^5  - (2 \pi)^{-2} \eps_{\mu \nu \rho \s}
\mbox{Tr}\,[ A^\nu
\partial^\rho A^\s - (2/3)  i A^\nu A^\rho A^\s ] \eqq j_\mu^5 + K_\mu^5,$$
where $j_\mu^5$ is the
gauge invariant point splitting regularized fermion current $\overline{\psi}
\g_\mu \g_5 \psi$. The current $j_\mu^5$ is not conserved because of the
anomaly, which is equivalent to the conservation of $J_\mu^5$.

In the usual discussion of the $U(1)$ problem (see e.g.~\cite{C}), the  current
$J_\mu^5$ has been discarded on the blame of its gauge dependence and the lack
of conservation of $j_\mu^5$ has been taken as the evidence that the axial
$U(1)$ is not a symmetry of the field algebra and therefore the problem of its
spontaneous breaking does no longer exist. Such a conclusion would imply that
time independent $U(1)$ axial transformations cannot be defined on the field
algebra $\F$ and not even  on its observable subalgebra $\F_{obs}$, which
contains the relevant order parameter. However, as argued by Bardeen on the
basis of perturbative renormalization (in local gauges), {\em the axial $U(1)$
transformations define a time independent symmetry of the  field algebra and of
its observable subalgebra}. This also follows from the conservation of
$J_\mu^5$ (equivalent to the anomaly of $j_\mu^5$), since in local
renormalizable gauges $J_\mu^5$ is a local operator, so that the standard
argument (see Sect.\,2) applies, i.e. eqs.\,(2.1),(2.2) hold. This implies that
(at least at the infinitesimal level) the r.h.s. of eq.\,(2.1) defines in this
case a symmetry of the field algebra  and in particular of the gauge invariant
observable subalgebra $\F_{obs}$, and the latter property is clearly
independent of the gauge. Therefore, there is no logical reason for a priori
rejecting the use of the gauge dependent current $J_\mu^5$ and of its
associated  Ward identities; one should only  keep in mind that in physical
gauges $J_\mu^5$ is a non local function of the observable (gauge independent)
fields.

The existence of axial $U(1)$ transformations of the observable subalgebra
$\F_{obs}$ implies that {\em the absence of parity doublets is a problem of
spontaneous symmetry breaking} and {\em the absence of massless Goldstone
bosons is reduced to the discussion  of local generation of the symmetry},
eqs.\,(2.1), (2.2), as in the case of the Higgs phenomenon.

In the local (renormalizable) gauges the time independent $U(1)$ axial symmetry
is generated by $J^5_\mu$ (and not by $j^5_\mu$) and the problem of massless
Goldstone modes does not arise because, as indicated by the perturbative
expansion and also by the Schwinger model~\cite{MPS}, {\em the correlation
functions of the (local) field algebra $\F$ are axial $U(1)$ invariant}.
However, the invariance of the vacuum functional $\Psio$, which defines the
local gauge quantization, does not mean that the symmetry is unbroken in the
irreducible representation of the observable  subalgebra $\F_{obs}$. In fact,
$\Psio$ gives a reducible representation of $\F_{obs}$ (as signaled by the
failure of the cluster property by the corresponding vacuum expectations), with
a non trivial center which is  generated by the large gauge transformations
$T_n$ and   is not pointwise invariant under $U(1)$ axial
transformations~\cite{MPS, QFT}. Thus, {\em the symmetry is broken in each pure
physical phase} ({\em $\theta$-vacuum sectors}) obtained by the diagonalization
of the $T_n$ (in the technical terminology by a {\em central decomposition} of
the observables) in the subspace $\F_{obs}\,\Psio$. It should be stressed that
the so obtained (gauge invariant) $\theta$-vacua do {\em not} provide well
defined representations of the field algebra $\F$, since the latter transforms
non trivially under $T_n$. This is at the origin of the difficulties (and
paradoxes) arising in the discussion of the chiral Ward identities
(corresponding to the conservation of $J_\mu^5$) in $\theta$-vacua
expectations.~\cite{CR} In the $\theta$ sectors a conserved axial current may
be constructed as a non local operator, typically by using for $J_\mu^5$ its
(non local) expression in terms of the observable fields in a  physical gauge.
The above discussion, in particular the lack of time independence in eq.\,(2.1)
as a consequence of the failure of relative locality between the current and
the order parameter, applies to such non local currents.

The resulting  mechanism for  the solution of the $U(1)$ problem can be made
explicit in the Coulomb gauge. In the Schwinger model, in  the Coulomb gauge
one has $K_0 = (e/\pi) A_1 = 0$, $K_1 = (e/\pi) A_0 $, so that $J_0^5 = j_0^5$
and the ($\theta$-)vacuum expectations of the  commutators $[ J_0^5(f_R,
t),\,A\,]$, $[\,j_0^5(f_R, t), A\,]$, $A \in \F_{obs}$, coincide and describe
the same mass spectrum; however, the time dependence in the limit $R \ra
\infty$,  in the first case can be ascribed to the non locality of the
conserved axial current, whereas in the second case it reflects the non
conservation of $j_\mu^5$.

\end{document}